\documentclass[final,5p,times]{elsarticle}
\usepackage{lineno}
\usepackage{url}
\usepackage{graphicx}
\usepackage{epstopdf}
\usepackage{amstext,amsmath,amssymb,amsthm}
\usepackage{upgreek}
\usepackage{enumerate}
\usepackage{txfonts}
\usepackage{subfigure}
\usepackage{multirow}
\usepackage{diagbox}
\usepackage{color,soul}
\biboptions{sort&compress}
\journal{Journal of High Energy Astrophysics}

\begin{document}

\makeatletter
\def\@cite#1{[{#1}]}
\makeatother

\begin{frontmatter}

\title{Design and Calibration of the High Energy Particle Monitor onboard the \textit{Insight}-HXMT}

\author[IHEP]{Xuefeng Lu}
\author[IHEP]{Congzhan Liu\corref{cor1}}\ead{liucz@ihep.ac.cn}
\author[IHEP]{Xiaobo Li}
\author[IHEP]{Yifei Zhang}
\author[IHEP]{Zhengwei Li}
\author[IHEP]{Aimei Zhang}
\author[IHEP,UCAS]{Shuang-Nan Zhang}
\author[IHEP]{Shu Zhang}
\author[IHEP]{Gang Li}
\author[IHEP]{Xufang Li}
\author[IHEP]{Fangjun Lu}
\author[IHEP,UCAS]{Yupeng Xu}
\author[IHEP]{Zhi Chang}
\author[IHEP]{Fan Zhang}
\cortext[cor1]{Corresponding author. }
\address[IHEP]{Key Laboratory of Particle Astrophysics, Institute of High Energy Physics, Chinese Academy of Sciences, Beijing 100049, China}
\address[UCAS]{University of Chinese Academy of Sciences, Chinese Academy of Sciences, Beijing 100049, China}
\begin{abstract}
Three high energy particle monitors (HPMs) employed onboard the Hard X-ray Modulation Telescope (\textit{Insight}-HXMT) can detect the charged particles from South Atlantic Anomaly (SAA) and hence provide the alert trigger for switch-on/off of the main detectors. Here a typical design of HPM with high stability and reliability is adopted by taking a plastic scintillator coupled with a small photomultiplier tube (PMT). The window threshold of HPM is designed as 1 MeV and 20 MeV for the incident electron and proton, respectively. Before the launch of \textit{Insight}-HXMT, we performed in details the ground calibration of HPM.  The measured energy response and its dependence on  temperature are taken as essential input of Geant4 simulation for estimating the HPM count rate given with an incident particle energy spectrum. This serves as a guidance for choosing a reasonable working range of the PMT high voltage once the real SAA count rate is measured by HPM in orbit. So far the three HPMs have been working in orbit for more than two years. Apart from providing reliable alert trigger, the HPMs data are used as well to map the SAA region.
\end{abstract}


\end{frontmatter}


\section{Introduction}
The Hard X-ray Modulation Telescope (HXMT), dubbed as \textit{Insight}-HXMT, was launched on 15th June 2017 to an orbit with an altitude of 550 kilometers and an inclination of 43 degrees. It is mainly composed of three kinds of collimating telescopes that work together to detect X/gamma rays in 1-250  keV \cite{Insight2019,Liu,Zhang2017,Chen,Xuelei}. The High Energy X-ray Telescope (HE covers 20-250 keV), is a main payload, designed with eighteen NaI(Tl)/CsI(Na) phoswich scintillators coupled with photomultiplier tubes (PMTs). As is well known, overmuch fluorescent lights irradiated by charged particles in scintillator will severely degrade the PMT performance because of saturation and nonlinearity, and can even shorten its lifetime. Therefore, it is very important to protect the HE from the damage induced by SAA, via switching it off according to the alert triggers provided by HPMs. On the other hand, it is also necessary to turn on the power supply immediately after exiting the SAA to obtain the maximum observational time. Although the SAA map has been measured many times by other satellites, it is still hard to trace the SAA boundary due to its evolution and drift \cite{Badhwar1997}. Therefore, charged particle monitors like these onboard RXTE, BeppoSAX \cite{Rothschild1997a,Campana2014}, Astrosat \cite{Rao2017a}, and as well the HPMs of \textit{Insight}-HXMT, are adopted specifically for measuring the live SAA boundary.

\begin{figure}[htb]
\begin{center}
\includegraphics[width=7cm]{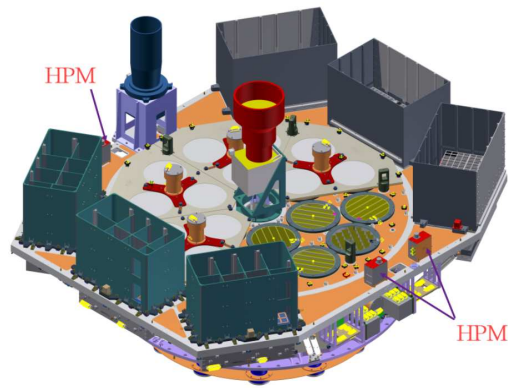}
\caption{HPM positions on the satellite platform.}
\label{platform}
\end{center}
\end{figure}

Similar to those on BeppoSAX and HEXTE, each HPM adopts the classical design for high stability and reliability, which takes a small plastic scintillator coupled with a photomultiplier tube. HPMs monitor the in-orbit particle flux  and transmit the pulse signal to electronic control system for count rate statistics. When the count rate exceeds a preset threshold, the electronic control system will be triggered, and then shut down the high voltage of the PMTs of the main detectors to protect them from damage. The window threshold of the HPM is designed as 1 MeV and 20 MeV for electron and proton, respectively. Benefited by the small size scintillator, the HPM maximum average count rate is expected to be less than 4000 events per second, according to the SAA electron and proton differential energy spectra available at SPENVIS website. Accordingly, the induced average anode current of the PMT shall be much less than 1 micro Ampere, which is safe for a HPM \cite{PMTmannual}. In addition, there are three HPMs backing up each other as shown in Figure \ref{platform}.

In this paper, firstly the mechanic and electronic design of the HPM are described in details. Then we introduce the ground calibration of the HPM with radioactive isotopes, part of which is the temperature response of the HPM in the range from --30 to +20 degrees. These in turn serve as the essential input of GEANT4 simulation to obtain the HPMs' detection efficiency for estimating the in-orbit count rate. Finally the relation between HPM count rate and PMT high voltage is derived and how it guides in-orbit PMT high voltage set is described. As a by-product, a map to outline the SAA region is derived as well. The in-flight performances of the HPM are given at the last.

\section{Detector design}
\begin{figure}[htb]
\begin{center}
\includegraphics[width=5cm]{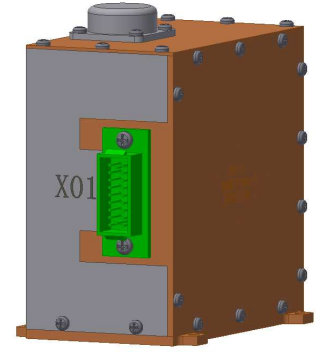}
\caption{Overall view of HPM.}
\label{shell}
\end{center}
\end{figure}

The HPM, shown in Figure~\ref{shell}, is mainly composed of a plastic scintillator coupled with a PMT, a divider PCB, a preamplifier PCB, a high voltage module, an aluminum shell and a connector.

\begin{figure}[htb]
\begin{center}
\includegraphics[width=5cm]{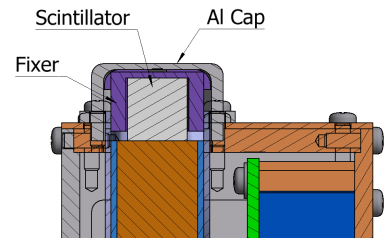}
\caption{Installation structure of plastic scintillator.}
\label{scintillator}
\end{center}
\end{figure}

The plastic scintillator BC440M produced by Saint-Gobain is sensitive to charged particles. Since a count rate of about 1 cps is expected outside the SAA, the detecting area of about 1 square centimeter is recommended according to the experience of the particle monitors on HEXTE and BeppoSAX. Here, we choose a cylinder scintillator with a diameter of 10 mm and a height of 10 mm. An aluminum cap with a thickness of 1 mm is used to limit the thresholds of electron and proton to 1 MeV and 20 MeV respectively. The scintillator, except for the surface coupled with PMT, is covered in sequence by reflective paint BC620 and Teflon to obtain higher light collection efficiency. The scintillator is adhesively coupled through a thin layer of transparent two-component-cured silicone rubber with a PMT R647-1 whose pins are welded on the divider PCB. The silicone rubber fixer shown in Figure~\ref{scintillator} houses the scintillator in the center of the PMT cathode, and absorbs vibrations. A 2 mm thickness magnetic shield tube is installed around the PMT to shield the geomagnetic field. The PMT cathode is about 5 mm lower than the magnetic shield tube to avoid the edge effect. Experiment shows that the magnetic intensity in the center of the magnetic shield tube is about one thousandth of the outside value. In order to absorb the vibrations during launch, a silicone rubber sleeve is inserted between the magnetic shield tube and PMT. As the magnetic shield tube and divider PCB are fixed by screws to the bottom plate of the HPM shell, the PMT is installed perpendicular to the divider PCB and kept upward.

The divider is designed based on the principle of equipartition of voltage, recommended by the PMT datasheet. A high voltage module S9100 from SITAEL is used to supply --1250 to 0 V for the divider (HPMs adopt negtive high voltage, for clarity and simplicity, the PMT high voltage will be expressed as positive).
It is covered by a polyamide plate and vertically fixed on the side aluminum plate by screws.

The electrons yielded by PMT are collected by a charge sensitive amplifier followed by a RC filter and a main amplifier. Since the decay time of the plastic scintillator is a few nanoseconds, the electron pulse from the PMT is about tens of nanoseconds contributed mainly by the transit time spread. Therefore, an integration time of hundreds nanoseconds is enough to collect those electrons completely and obtain the maximum pulse height. The pulse width is about 450 ns for normal events. This allows the HPM to detect easily high-flux charged particles in SAA without saturation. In addition, a differential output is adopted to reduce the common mode interference signals. The electronic noise is about 10 mV. These components are placed on another PCB fixed to the side aluminum plate.

\begin{figure}[htb]
\begin{center}
\includegraphics[width=5cm]{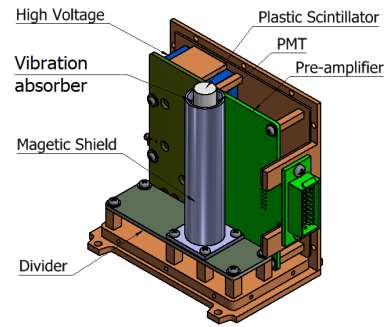}
\caption{Internal structure diagram of HPM.}
\label{structure}
\end{center}
\end{figure}

The HPM shell is composed of five mechanical structural components screwed together. A mechanical stop design is adopted for each plate to keep external light and electric field out of the detector. The HPM outline dimension is 111 mm $\times$ 64 mm $\times$ 104 mm. The detector design is very compact and has a high stability. Ground experiments have shown that the HPM can withstand vibrations with 500 times the gravitational acceleration and work well from --30 to +20 degrees.

When a charged particle deposits energy in the scintillator, a voltage pulse with a height proportional to the deposited energy will be generated as the output of HPM. The voltage pulse is then sent to a comparator in the electronic control system through a two-meter long cable. The comparator will also receive another DC reference voltage given by the electronic control system. The initial DC voltage is 200 mV and can be programmed by tele-commands. When the pulse amplitude exceeds this value, the comparator will give a positive square wave pulse to trigger the counter. If the count rate exceeds a certain value which is set by tele-commands to 10 cps within three seconds, the satellite is supposed to enter the SAA region and, accordingly, the high voltage of the PMTs in the HE detectors will be turned off. Vice verse, the detectors will be turned on once the satellite moves out of SAA region.

\section{Ground calibrations and simulations}
The HPM is regarded only as a counter of charged particles, and the pulse height of each event is not measured by the electronic control system. Ground calibrations were carried out to characterize the HPM and a Geant4 simulation was also made to investigate the responses of the HPM to electrons and protons. Based on these data, a relation between the PMT high voltage of the HPM and the expected average count rate in SAA is built carefully, which in turn provide reference for adjusting the PMT high voltage of the HPM in orbit.

\subsection{Relation between detection threshold and PMT high voltage}
In principle, when a particle deposits energy in plastic scintillator through ionization, a certain amount of fluorescent photons will be produced. Part of them will convert to electrons at the PMT cathode because of  photoelectric effect. Then those electrons are driven by an electric field toward the first dynode, where more new electrons will be excited by collision and further driven toward the next dynode. By multi-amplifying, the PMT anode will eventually output a pulse with numerous electrons, which will be collected by a charge sensitive amplifier and shaped to be a voltage pulse with a height ranging from hundreds millivolts to several volts.

According to the literature \cite{Schmidt2002,Mukhopadhyay}, the yield of light is almost linear to the deposited energy in plastic scintillator. Therefore, given a fixed collection efficiency and quantum efficiency, the number of electrons reaching the first dynode is also linearly related to the deposited energy. On the other hand, the total gain \textit{g} of the PMT dynodes has a power law relation to the PMT high voltage \textit{V} at the equipartition of voltage, shown as \textit{g} $\propto$ \textit{V$^{kn}$} in the PMT handbook \cite{PMTmannual2}, where \textit{n} is the number of dynodes and \textit{k} is a constant value determined by the material characteristics of dynode. Furthermore, the electronic amplifier is a linear system whose contribution to the pulse height can be regarded as a constant factor. Therefore, the pulse height \textit{H} is a function of the deposited energy \textit{E} and the PMT high voltage \textit{V}, and can be expressed as
\begin{equation}
\label{eqb}
H=a\times E\times V^c+b,
\end{equation}
where \textit{a} is the joint constant coefficient, \textit{c} = \textit{kn}, and \textit{b} is the baseline of the electronic system. Once those three parameters are determined by experiment, the relation between the detection threshold and the PMT high voltage will be built.

\begin{figure}[htbp]
\begin{center}
\includegraphics[width=8cm]{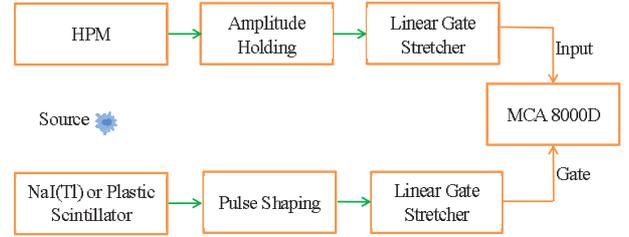}
\caption{Diagram of calibration experiment setup.}
\label{diagram}
\end{center}
\end{figure}

To figure out Eq.~\ref{eqb}, ground calibration was carried out with radioactive sources ${}^{241}$Am, ${}^{22}$Na and ${}^{137}$Cs. The calibration experiment setup diagram is shown in Figure~\ref{diagram}. A plastic scintillator embedded with an ${}^{241}$Am source supplies the alpha signal for coincidence detection of 59.5~keV gamma ray signal from the HPM. For the radioactive source ${}^{22}$Na, gamma rays pair of 511.0 keV are born out of the positron-electron annihilation in the opposite direction. Once one gamma-ray is caught by a NaI(Tl) detector CH132-06 made by HAMAMATSU, the other gamma-ray photon detected by HPM can be found out coincidently. The pulse output from the plastic or NaI(Tl) scintillation detector is shaped and stretched into a rectangle shape with a width of 5 $\mu$s and an amplitude of 5 V. Similar story happens to HPM but with the amplitude kept the same as the original pulse height. HPM detects photons with mono-energy of  661.7 keV and K-edge line with an average energy of 32.9 keV from the radioactive source ${}^{137}$Cs.

\begin{figure}[htb]
\begin{center}
\includegraphics[width=8cm]{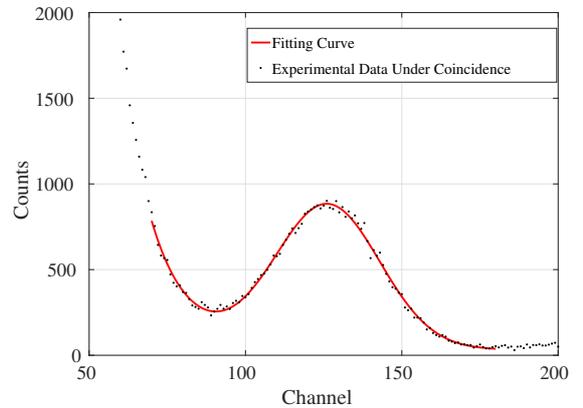}
\caption{Full-energy peak spectrum of 59.5 keV gamma ray}
\label{spec1}
\end{center}
\end{figure}

\begin{figure}[htb]
\begin{center}
\includegraphics[width=8cm]{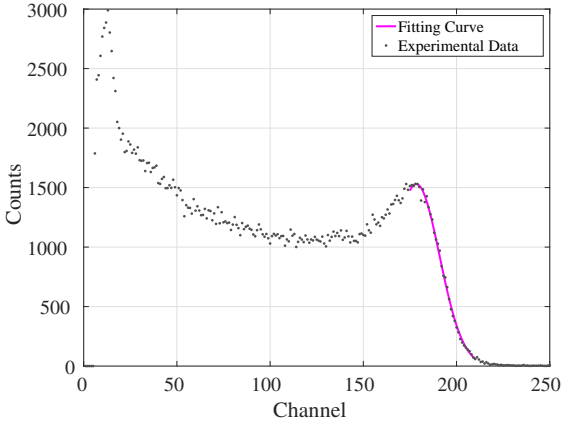}
\caption{Compton spectrum of 661.7 keV gamma ray}
\label{spec2}
\end{center}
\end{figure}

\begin{figure}[htb]
\begin{center}
\includegraphics[width=8cm]{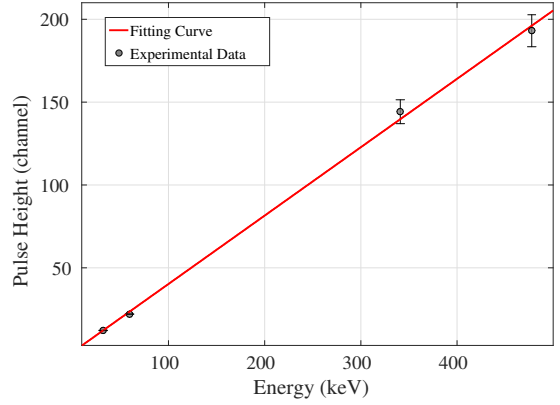}
\caption{Energy response at PMT high voltage 687.5 V. }
\label{EC}
\end{center}
\end{figure}

The energy response for the 59.5 keV and 661.7 keV incident photons at two different PMT high voltages are shown in Figure~\ref{spec1} and Figure~\ref{spec2}, respectively. The full-energy peak and Compton peak are extracted by Gaussian fit first, and then the real Compton scattering edge positions of 340.7 keV (corresponding to 511.0 keV) and 477.3 keV (661.7 keV) are calculated with the method in reference \cite{G.C-1973} and the error estimation in \cite{Dietze1982}. As shown in Figure~\ref{EC}, the pulse height is almost linear to the deposited energy, which is consistent with previous studies in \cite{Schmidt2002,Mukhopadhyay}. The relation at the PMT high voltage 687.5 V (corresponding to the PMT control voltage 2.2 V) can be expressed as
\begin{equation}
\label{eq1}
H(E)\vert_{\rm2.2~ V}=0.4127\times\textit{E}-1.028,
\end{equation}
where \textit{H}(\textit{E}) stands for the pulse height, \textit{E} the deposited energy and constant --1.028 the baseline (1.028 stands for about 10 mV).

\begin{figure}[htb]
\begin{center}
\includegraphics[width=8cm]{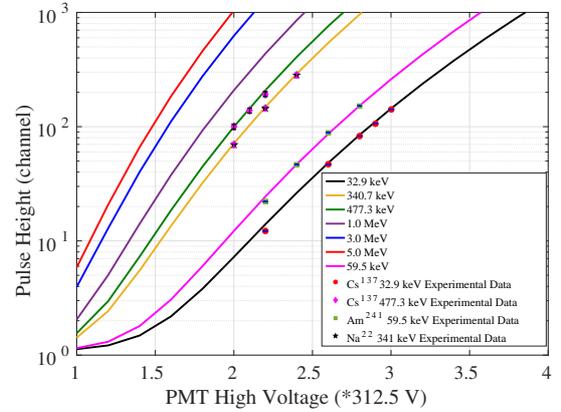}
\caption{High voltage versus pulse height corresponding to different deposited energy. }
\label{map}
\end{center}
\end{figure}

In addition, the full-energy peak corresponding to an incident energy of e.g. 59.5 keV manifests itself with a specific relation with the PMT high voltage (see Figure.~\ref{map}). Such a relation can be well fitted with a power law function \cite{PMTmannual2} and an example for the incident 59.5 keV photons is shown as
\begin{equation}
\label{eq2}
H(V)\vert_{\rm59.5~keV}=0.05103\times \textit{V}^{7.766}-1.028,
\end{equation}
\textit{V} is the PMT high voltage divided by 312.5. Based on the above equations, the relation between the pulse height and the high voltage at different deposited energies can be expressed as
\begin{equation}
\label{eq3}
H(E,V)=8.57\times10^{-4}\times \textit{E}\times \textit{V}^{7.766}-1.028.
\end{equation}
Asuming an electronic threshold of \textit{H}$_{\rm{th}}$, the detection threshold \textit{E}$_{\rm{th}}$ at different PMT high voltage is given as
\begin{equation}
\label{eq4}
E_{\rm{th}}=\frac{H_{\rm{th}}+1.028}{8.57\times10^{-4}\times V^{7.766}}.
\end{equation}
This relation holds for the deposited energies ranging from several keV to MeV. Due to linearity of the energy response, such an energy band is sufficient for us to calculate the HPM count rate in SAA.

\subsection{Temperature response}

\begin{figure}[htb]
\begin{center}
\includegraphics[width=8cm]{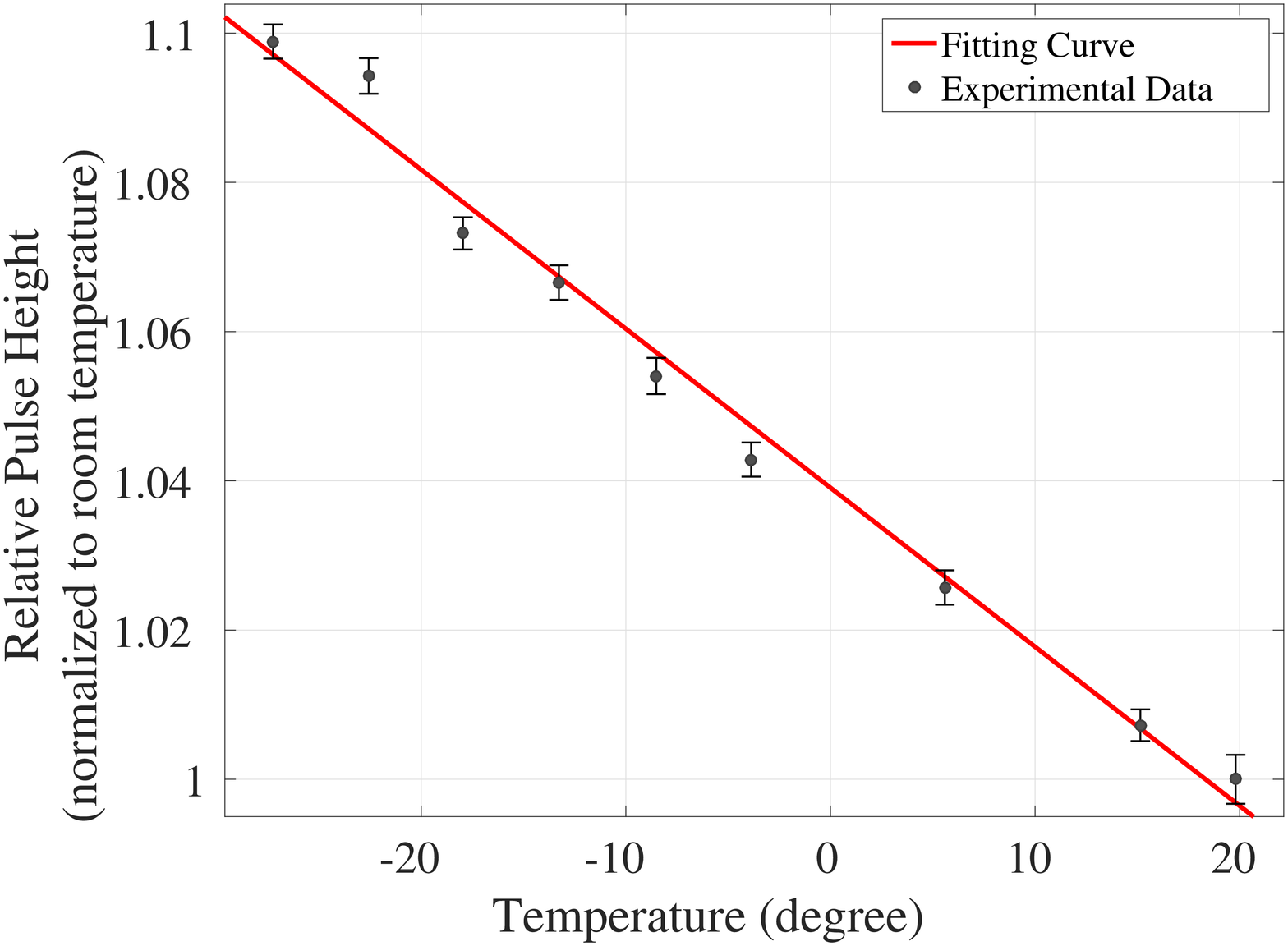}
\caption{Full-energy peak of 59.5 keV gamma rays at different temperature. }
\label{Temp}
\end{center}
\end{figure}

Due to the influence of low temperature environment in space, the HPM will work at the temperature far below 20 degrees. This will remarkably change the HPM detection threshold obtained at room temperature, as the quantum efficiency of PMT photocathode is sensitive to temperature. At laboratory, temperature response of a HPM has been investigated in the range from --30 to +20 degrees, equipped with 59.5 keV incident photons irradiated by an ${}^{241}$Am source. It is obvious in Figure~\ref{Temp} that the pulse height almost linearly increases with decrease of the temperature. It turns out that a variability of 10\% for HPM gain can enclose the entire range of the temperature experienced by HPM in this calibration experiment. The linear fitting of the experimental data results in a slope of --0.21\% per degree,  which is accounted for by multiplying in Eq.~\ref{eq4} a coefficient of 1--0.21\%(\textit{T}--20) to the item 8.57$\times$10$^{-4}$ when the HPM works at temperature \textit{T}.

\begin{figure}[htb]
\begin{center}
\includegraphics[width=8cm]{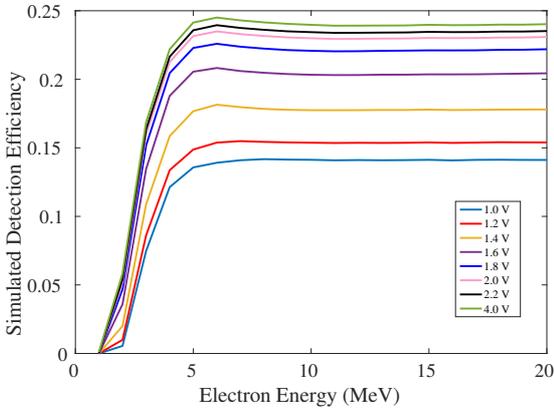}
\caption{Simulated detection efficiency for electron at different high voltage. }
\label{eff}
\end{center}
\end{figure}

\begin{figure}[htb]
\begin{center}
\includegraphics[width=8cm]{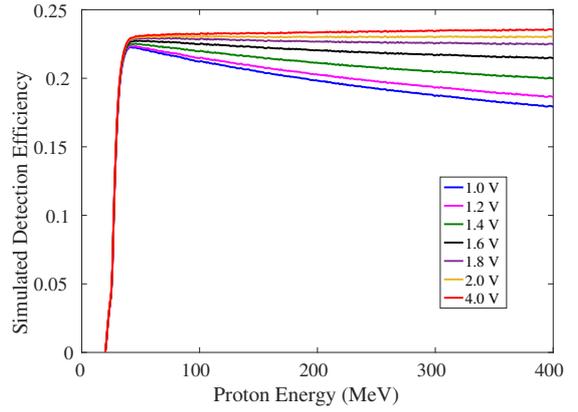}
\caption{Simulated detection efficiency for proton at different high voltage. }
\label{pronton}
\end{center}
\end{figure}

\begin{figure}[htb]
\begin{center}
\includegraphics[width=8cm]{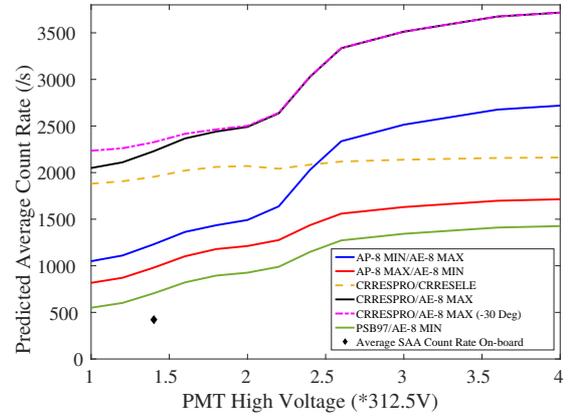}
\caption{Average count rate in SAA predicted on the basis of different models. The default flux threshold for each models is 1 cm$^{-2}$ s$^{-1}$. The trajectory duration time is 10 days. The CRRESPRO version is quiet, and the CRRESELE version is Ap 25 -- 55. The rosy dotted line is the result at the ¨C- 30 degrees, while others are predicted at room temperature. The diamond point shows an actual measurement result of a HPM. }
\label{predict}
\end{center}
\end{figure}

\subsection{Detection efficiency simulated by Geant4}

Since it is usually hard to measure the absolute detection efficiency to electrons and protons for HPM through the ground-based experiment, people usually turns to Geant4 tool for rough estimation via simulation. Here in Geant4 the mass model of HPM comes from the flight payload, and the physical processes cover ionization, multiple scattering, bremsstrahlung, Compton scattering, gamma conversion, and positron-electron annihilation. The setups of the simulation are what follows:
electron and proton sampled isotropically in 2$\pi$ direction at the surface of the aluminum cap; energetic electron in the range 1-20 MeV, proton in the range 20-400 MeV. The incident charged particle will be detected by HPM once its deposited energy beyond the threshold shown in  Eq.~\ref{eq4}. Here the detection efficiency is defined as the ratio of the number of detected particles to that of the total incident events. Figure~\ref{eff} shows the detection efficiency of electrons at different high voltages. Electron below 1 MeV can not penetrate the aluminum cap and the corresponding efficiency is zero. The detection efficiency increases with energy and reaches a plateau of roughly 24\% at energies above 5 MeV under a maximum high voltage of 1250 V. This efficiency is  comparable to ratio of the surface area of the scintillator to that of the aluminum cap. Similar behavior presents as well for protons and the results are shown in Figure~\ref{pronton}. The maximum detection efficiency of proton above 40 MeV is about 23\%, slightly lower than that of electron probably due to relatively less energy deposition of proton. Further increase in proton energy will lead to even less energy deposition and hence smaller detection efficiency.

\subsection{Results and discussions}
SPENVIS (ESA's SPace ENVironment Information System) is a website interface to models of the space environment and its effects; including cosmic rays, natural radiation belts, solar energetic particles and so on. There are many models on SPENVIS to depict the trapped particle radiation belts. With the input of the satellite orbital altitude, inclination angle, trajectory duration time and other information, the geographical position will be firstly generated. And the proton and electron spectra in the corresponding orbits can be given by different radiation belt models implemented on SPENIVIS website. For low earth orbit satellite \textit{Insight}-HXMT, only AP-8/AE-8, AP-9/AE-9, CRRESPRO/CRRESELE and PSB97 are suitable. In this paper, the trapped proton and electron spectra along the orbit from those models are convolved with the HPM detection efficiency to obtain the SAA map of each model, and then the average count rate is obtained from the predicted SAA map using the same counting threshold of 10 counts per second. The AP-9 and AE-9 models are not used as they cannot accurately predict the SAA zone for lack of the real time differential particle spectra along the orbit. The NASA AP-8 and AE-8 radiation belt models are still the de facto standards for engineering applications. This is mainly due to the fact that up to now they are the only models that completely cover the region of the radiation belts, and have a wide energy range for both protons and electrons. Centered on AP-8 and AE-8 models, we choose five combined models to give the worst and the best prediction. As shown in Figure~\ref{predict}, CRRESPRO/CRRESELE model gives a relative non-accurate prediction for lack of electron data in SAA and a narrow proton band. The combination CRRESPRO/AE-8 MAX gives the highest prediction and the PSB97/AE-8 MIN gives the lowest value. The dot-dashed line obtained at - 30 degrees only shows obviously influence on count rate measured at low PMT high voltage. The increase in count rate due to a decrease in operating temperature will not have a particularly significant effect on the average count rate level. All the predictions give average count rate no more than 4000 events per second, while the low count rate section between 312.5V and 625 V is the most beneficial for HPM working. As the count rate of the HPM in the non-SAA region is rather small, which has a typical value of 1 cps, it is sufficient to judge whether the HPM is within SAA with a count rate beyond hundreds. Correspondingly, the PMT high voltage is recommended to be between 312.5 V and 625 V.

\section{Status in orbit}

\begin{figure}[htb]
\begin{center}
\includegraphics[width=8cm]{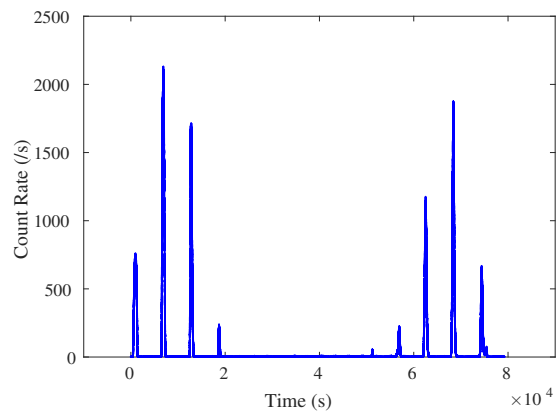}
\caption{Real time count rate of one HPM. Zero time in the figure is 2018-09-01T02:00:00.000 UTC. }
\label{count}
\end{center}
\end{figure}

\begin{figure}[htb]
\begin{center}
\includegraphics[width=8cm]{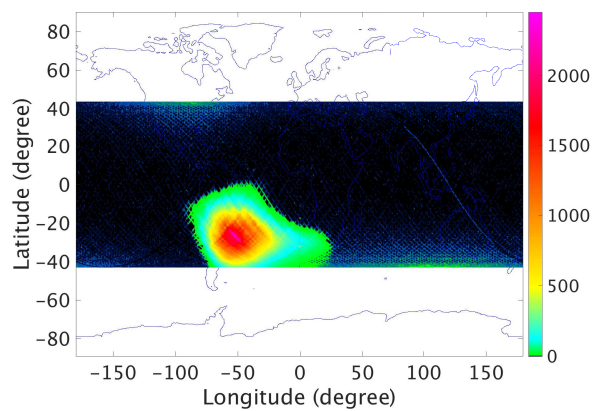}
\caption{SAA map from HPM data.}
\label{SAA}
\end{center}
\end{figure}

After entering the orbit, the HPMs were powered on, and the PMT high voltage was set to  437.5 V which is in the recommended range. The HPM surface temperature varies between --18 and --22 degrees during flight. Figure~\ref{count} shows the real count rate measured by an HPM over a specific period of time. The pulses indicate that the HPM travels through the SAA region occasionally. A contour map of the in-orbit HPM count rate is shown in Figure~\ref{SAA}, from which the SAA region is clearly outlined. The HPM detections show that, the count rate is roughly 1  in most area outside SAA and thousands in the central region of
the SAA. The average count rate of the HPM in SAA is about 424 cps, which is marked in Figure~\ref{predict} for comparison. The deviation of this value from the model prediction may due to that the environment chosen by the model is slightly different from the real one.

\section{Conclusions}
As a compact and reliable particle monitor, \textit{Insight}-HXMT HPM went through a series of ground tests and is currently working smoothly in orbit. The ground calibrations provide a reliable reference for adjusting the in-orbit high voltage and detection threshold. With the recommended PMT high voltage, apart from providing reliable normal alert trigger for all \textit{Insight}-HXMT payloads, HPMs' results are used as well to map the SAA region. This method can also be applied to the  particle monitors with a similar design onboard the future space-born telescopes. Along with the service of \textit{Insight}-HXMT in orbit, HPM will accumulate abundant data to help us understand the radiation environment of low earth orbit .

\section*{Acknowledgment}\label{sect:acknowledgement}

This work was supported by the National Key R\&D Program of China (2016YFA0400800, 2016YFF0200802), the Strategic Priority Research Program of the Chinese Academy of Sciences (Grant No. XDB23040400) and the National Natural Science Foundation of China under grant U1838104, U1838201, U1838202.

\bibliographystyle{model1a-num-names}
\bibliography{HPMdesign}



\end{document}